\documentstyle[11pt,newpasp,twoside,psfig]{article}
\def\gapprox{\lower.4ex\hbox{$\;\buildrel >\over{\scriptstyle\sim}\;$}}
\def\lapprox{\lower.4ex\hbox{$\;\buildrel <\over{\scriptstyle\sim}\;$}}

\reversemarginpar

\begin{document}
\title{Pulse emission mechanisms}
 \author{Don Melrose}
\affil{School of Physics, University
 of Sydney, Sydney NSW 2006 Australia}

\begin{abstract}
High-energy and radio emission mechanisms for pulsars are reviewed. The source region for high energy emission remains uncertain. Two preferred radio emission mechanism are identified. Some difficulties may be resolved by appealing to nonstationary pair creation distributed widely in height.

\end{abstract}

\section{Introduction}

Collectively, pulsars emit over essentially the entire electromagnetic spectrum. However, most pulsars are observed only in a relatively narrow range, $\sim100\rm\,MHz$--$\sim10\rm\,GHz$. The basic properties determined for most pulsars are the period, $P$, and the period derivative, $\dot P$, which determine the magnetic field, $B\propto(P{\dot P})^{1/2}$, the characteristic age $P/2{\dot P}$ and the spin-down power $\propto{\dot P}/P^3$. Also available for most pulsars are the integrated pulse profile, the position angle of the linear polarization, which determines the inclination angle between the magnetic and rotational axes, and the dispersion measure, which provides an estimate of the distance to the pulsar. Based on the location in the $P$--$\dot P$ plane, pulsars are classified as young ($\lapprox10^5\rm\,yr$), old, recycled (or millisecond), or magnetar. It is remarkable that despite the large ranges in $P$, $B$, the variations in pulse profile between classes are similar to those within classes. This suggests that the radio emission mechanism cannot be sensitive to either $P$ or $B$. The source of the radio emission appears to be between inner and outer gaps, seemingly far from these sites where parallel electric fields are thought to accelerate primary particles and trigger pair cascades. Only a small fraction of radio pulsars are also observed at high energies, in the X-ray and gamma-ray ranges, which emissions should provide a direct signature of particle acceleration and pair creation. However, there remains a major uncertainty in the location of the source region, with both inner-gap and outer-gap models considered viable. 

High-energy emission mechanism are summarized in section~2 and radio emission mechanisms are summarized in section~3. The identification of source regions is discussed in section~4.

\section{High-energy emission mechanisms}
The high-energy emission mechanisms that need to be considered for  pulsars are synchrotron emission, curvature emission and inverse Compton emission (IC), all of which are relatively well known. A less familiar, resonant form of IC emission (RIC) is important.

A preliminary point is that in a relativistic quantum treatment, the energy of an electron has discrete values $\varepsilon_n=(m^2c^4+p_z^2c^2+2neBc^2\hbar)^{1/2}$ where $p_z$ is the component of the momentum along the field lines, and $n=0,1,\ldots$ is the Landau quantum number. In a pulsar the lifetime of the excited states is very short, and all electrons quickly relax to their ground state, $n=0$. 

Synchrotron emission occurs only if highly relativistic pairs are created with $n\gg1$, corresponding to $\gamma\sin\alpha\gg1$, $\gamma=\varepsilon_n/mc^2$, $\alpha=$ pitch angle. The emission is strongly concentrated around an angle $\theta=\alpha$, with a broad frequency spectrum peaked at $\omega\approx\omega_B\gamma^2\sin\theta$, where $\omega_B=eB/m$ is the cyclotron frequency. The power radiated is $(e^2\omega_B^2/6\pi\varepsilon_0c)\gamma^2\sin^2\theta$, and the radiation is highly linearly polarized, $\sim70\%$, in the direction perpendicular to the projection of the magnetic field on the plane of the sky. For a power-law energy spectrum of radiating particles, $N(\gamma)\propto\gamma^{-a}$, the intensity spectrum is a power law, $I(\omega)\propto\omega^{-(a+1)/2}$.

Curvature emission by an individual electron with Lorentz factor $\gamma\gg1$ moving along ($n=0$) a magnetic field line with radius of curvature $R_{\rm c}$ is strongly concentrated around $\theta=0$, with a broad frequency spectrum peaked at $\omega\approx\omega_{\rm c}\gamma^3$, $\omega_{\rm c}=c/R_{\rm c}$. The power radiated is $(e^2\omega_{\rm c}^2/6\pi\varepsilon_0c)\gamma^4$, and the radiation is highly linearly polarized in the direction perpendicular to the plane containing the curved magnetic field line. For a power-law energy spectrum of radiating particles the intensity spectrum is a power law with index $-(a+2)/3$.

IC emission is actually Thomson scattering by highly relativistic electrons. An initial photon with frequency $\omega_0$ is scattered into a final photon with frequency $\omega\sim\omega_0\gamma^2$ propagating nearly along the direction of the initial electron. The power radiated is $\approx\sigma_TcW_0\gamma^2$, where $\sigma_T=(8\pi/c)(e^2/4\pi\varepsilon_0mc^2)^2$ is the Thomson cross section and $W_0$ is the energy density in the initial photons. The polarization has no characteristic value in general, and the spectrum is synchrotron-like. Relativistic quantum effects sharply reduce the cross section, from the Thomson to the Klein-Nishina value, for $\omega_0\gg mc^2/\gamma\hbar$.

RIC emission is Thomson scattering by highly relativistic electrons in a magnetic field such that the wave frequency, $\omega_0$, is near the relativistic cyclotron frequency, $\omega_0\approx\omega_B/\gamma$. When this condition is satisfied, the Thomson cross section is greatly enhanced, by a factor $\sim\omega^2_0/(\omega^2_0-\omega_B^2/\gamma^2)$ (Canuto, Lodenquai \& Ruderman 1971) dependent on the polarization of the photon (Melrose \& Sy 1972). The maximum enhancement factor is $\sim\omega_B^2/\Gamma^2$, where $\Gamma$ is the width of the cyclotron resonance. RIC involves scattering in which the initial and final state of the electron is $n=0$ and the (virtual) intermediate state is $n=1$. The relevant $\Gamma$ is the inverse lifetime of the first excited state (Herold 1979), which is $\Gamma=e^2\omega_B^2/3\pi\varepsilon_0mc^3\gamma$. The importance of RIC in high-energy emission from pulsars was emphasized by Daugherty \& Harding (1989) and Dermer (1990). 

\section{Models for high-energy emission}

There are both inner-gap and outer-gap models for the high-energy emission, and there is no clear consensus on which type of model is to be preferred. 

High-energy emission should provide a signature of acceleration and pair production in inner-gap models, provided a small fraction of high-energy photons escape. Recent inner-gap models (Zhang \& Harding 2000, Hibschman \& Arons 2001, Arendt \& Eilek 2002) confirm the general features of earlier models, generally favoring a higher fraction of lower energy, $\gamma\lapprox100$, pairs and pair creation extending to greater heights than earlier models. The efficiency of conversion of the energy in primary particles can be very high, when acceleration is radiation-reaction limited due to curvature emission losses. RIC emission off thermal photons from the surface of the neutron star can be particularly effective (Zhang \& Harding 2000). Synchrotron emission from the secondary pairs can contribute to the observed high-energy emission, provided the magnetic field is not too strong. Existing models are not satisfactory for relatively strong fields for reasons discussed by Zhang \& Harding (2000). Another effect that has not been included is that for $B\gapprox0.1$--$0.2\,B_c$ ($B_c=4\times10^9\rm\,T$), the secondary pairs are created predominantly in their ground state (Weise \& Melrose 2002) or as bound pairs (Usov \& Melrose 1996), in which cases synchrotron emission is absent.

\begin{figure}
\centerline{\psfig{file=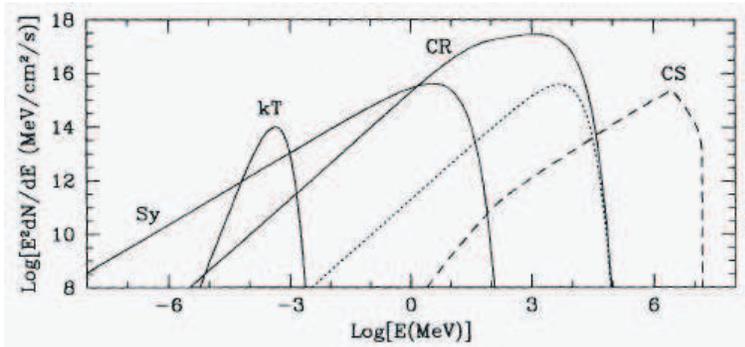,width=10cm}}
\caption{The energy spectrum due to synchrotron emission (Sy), thermal surface flux (kT), curvature emission (CR) and IC off Sy (CS) for a radiation-reaction limited spectrum with maximum energy $E_e=3\rm\,GeV$ and $B=3\times10^8\rm\,T$ (Romani 1996).}
\end{figure}

Outer-gap models (e.g, Cheng, Ho \& Ruderman 1986) involve particle acceleration, high-energy emission and pair production in an outer gap, relatively near the light cylinder. Some of the relevant parameters in the outer gap are quite different from the inner gap: $B$ is much weaker, $R_{\rm c}$ is much larger, and the density of thermal photons from the surface of the star is much lower. Emission of photons with a given energy due to synchrotron emission or curvature emission requires particles with a much higher $\gamma$ in the outer gap compared with the inner gap. The low density of thermal photons implies that they are unimportant as targets for IC emission. Romani (1996) compared the contributions to the high-energy emission due to the various processes for a particular choice of parameters, as illustrated in figure~1. Qualitative arguments in favor of outer-gap models are that they lead naturally to the double-peaked profiles of the form observed in some sources, and imply harder spectra from older, more efficient pulsars (Romani 1996). The sensitivity of the models to various parameters, notably $B$ and the inclination angle between rotation and magnetic axes, provides scope for accounting for most of the observed features of high-energy emission from individual pulsars.

The properties of inner-gap and outer-gap models (Cheng \& Zhang 1999) were compared by Zhang \& Harding (2000), concentrating on X-ray emission. Qualitatively, nonthermal X-ray emission in inner-gap models is attributed to RIC emission due to upward-propagating pairs, and in outer-gap models it is attributed to synchrotron emission by downward propagating pairs. Thermal X-ray emission from polar-cap heating is determined by the energy flux of secondary pairs incident on the polar cap, heating it, and this flux is different in the different models. However, these and other differences have yet to provide definitive arguments that distinguish between the two types of model.

The uncertainty as to whether the source region is in the inner-gap or the outer-gap, or even in the wind (Kirk, Sk\ae raasen \& Gallant 2002), is likely to be resolved only when more detailed observational data on the high-energy emission becomes available. A  hint as to how a partial resolution might arise is through an obscuring of the distinction between inner and outer gaps. Inefficient screening of the accelerating electric field (e.g., Hibschman \& Arons 2001) allows pair production and high-energy emission from an inner gap to extend to greater heights than previously thought, and the dependence of the location of the outer gap on the global current (Hirotani \& Shibata 2001) allows pair production and high-energy emission from an outer gap to occur at lower heights than previously thought. It may be that acceleration and pair production occur at a significant level over a wide range of heights.

\section{Radio emission mechanisms}

Amongst the many pulsar radio emission mechanisms that have been proposed, only those that involve some form of plasma instability or maser action are mentioned here. There are at least five distinct instability or maser mechanisms currently under consideration. The two most favorable of these are discussed here: a form of maser curvature emission and relativistic plasma emission (beam-driven Cerenkov instability). The other three are described more briefly.

\medskip
\noindent
{\it Maser curvature emission}: Although initially thought impossible (Blandford 1975; Melrose 1978), maser curvature emission is possible when either curvature-drift (Luo \& Melrose 1992) or field line torsion (Luo \& Melrose 1995) are taken into the account. There are two unfavorable features of curvature-drift-driven maser. First, it is sensitive to $B$, and so cannot plausibly be applied to all classes of pulsars. Second, maser emission is possible only for Lorentz factors above a relatively high threshold, $\gamma\gapprox10^4$. However, like all relevant instability mechanisms, growth requires that the particle distribution function be an increasing functions of Lorentz factor, $df(\gamma)/d\gamma>0$, and this condition is satisfied only for $\gamma\lapprox100$ in current models for pair cascades (Hibschman \& Arons 2001; Arendt \& Eilek 2002). The torsion-driven maser requires that the field line not be confined to a plane, so that it has two different radii of curvature in orthogonal directions. There are two possible sources of torsion. One is rotation, which causes a sweeping back of the field lines. This is significant only near the light cylinder. The other is a non-dipolar component of the field, which tends to produce torsion in the inner magnetosphere. For example, a quadrupolar component would provide torsion in general, with the ratio of the radii of curvature increasing linearly with radial distance from the center of the star. Like the curvature-drift-driven maser,  growth is possible only above a minimum $\gamma$, but in this case the threshold, $\gamma\gapprox40$ (Luo \& Melrose 1995), is compatible with $df(\gamma)/d\gamma>0$ from pair cascade models. 

A strong argument in favor of torsion-driven maser curvature emission is that the mechanism depends sensitively only on the primary radius of curvature, and is not particularly sensitive to the ratio of the radii of curvature. Hence the properties of the emission is determined by the dipolar component of the magnetic field, which is plausibly similar in all pulsars. As a consequence, the mechanism can account naturally for the remarkable similarity in the frequency range for radio emission from all classes of pulsars. Another argument in favor of the mechanism is that growth occurs only in one linear polarization (that orthogonal to the dominant polarization in incoherent curvature emission), and this is consistent with observational evidence on the plane of polarization (Radhakrishnan \& Deshpande 2001).

\medskip
\noindent
{\it Relativistic plasma emission}:
Some form of beam-driven wave growth has long been favored as the pulsar radio emission mechanism, but various difficulties with it have led to ongoing doubts about its viability. Generically, a beam-driven Cerenkov instability requires (a)~that the growing waves have refractive index greater than unity (phase speed  $<c$), and (b)~that the distribution of particles includes a counterstreaming motion such that one component is streaming through another at greater than the phase speed of the growing wave. Effective growth requires that the growth rate for the instability be greater than effective loss rates for the growing waves. Seemingly plausible beams are primary particles streaming through secondary pairs, or electrons streaming relative to positrons to provide the required current density, but for both the growth rate is far too small. One seems to be forced to appeal to some form of nonstationarity in the pair creation that results in the faster particles from a following bunch overtaking the slower particles in a preceding bunch (Usov 1987).

\begin{figure}
\centerline{\psfig{file=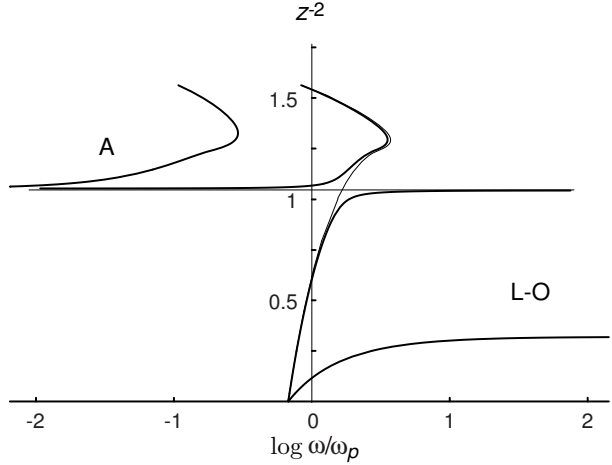,width=8cm}}
\caption{Dispersion curves, $\omega/\omega_p$ as a function of $z^{-2}=n^2\sin^2\theta$, for $\langle\gamma\rangle\sim2$, $v_A/c\sim5$ and $\theta=0$ (thin intersecting curves), $\theta=0.1$ (inner solid curves) and $\theta=1$ (outer solid curves) showing how the Alfv\'en (A) and L-O~modes separate with increasing $\theta$ (Melrose \& Gedalin 1999).}
\end{figure}

When the dispersive properties of waves in a pulsar plasma are considered in detail (e.g., Lyutikov, Blandford \& Machabeli 1999), further difficulties arise. In the early literature, a possible analogy with growth of Langmuir waves in solar radio bursts suggested that the growing waves might be longitudinal waves near $\omega_p/\gamma^{3/2}$ in a relativistically-streaming, monoenergetic model. However, the plasma is expected to have a large spread, $\langle\gamma\rangle$, in Lorentz factor (e.g., in the rest frame of the plasma). When such a spread is taken into account, parallel-propagating ($\theta=0$) longitudinal waves have phase speeds $<c$ only above a relatively high frequency $\sim\omega_p\langle\gamma\rangle^{1/2}$ in the plasma rest frame. This frequency is further boosted, by the Lorentz factor of the streaming motion, in the pulsar frame. Oblique (L-O mode) waves have phase speed $<c$ only for a tiny range of angles about $\theta=0$. The dispersion curves in the rest frame are plotted in figure~2 for a small value of $\langle\gamma\rangle$; beam-driven growth of L-O~mode waves is possible only for sufficiently small angles, $\theta\ll1/\langle\gamma\rangle$, for which the dispersion curve extends into the region above the line $z^{-2}=1$ in the figure.

These features lead to several unresolved problems (Melrose \& Gedalin 1999).  First, the frequency of the beam-driven waves (Doppler boosted to the pulsar frame) seems too high to account for all pulsar emission. Second, due to the curvature of the field lines, waves rapidly leave the tiny range of angles where growth is possible, severely restricting effective growth. The suggestion that beam-driven growth may occur in other wave modes in the pulsar plasma leads to other difficulties: the X~mode (in a non-gyrotropic plasma) does not couple directly to the beam; Alfv\'en waves can grow at lower frequencies, but cannot escape directly due to a stop band, cf., curve A in figure~2.

The required large growth rate in a highly relativistic plasma precludes a maser mechanism: a maser operates in the random-phase regime, which requires that the growth rate be less than the bandwidth of the growing waves, which is very small in a highly relativistic plasma. A (resonant) reactive or hydrodynamic version of the instability  applies for $\omega\sim\omega_p\langle\gamma\rangle^{1/2}$. There is also a nonresonant reactive instability, which causes waves in a beam mode to grow at lower frequencies. Appeal to this nonresonant instability partly alleviates some of the difficulties (Gedalin, Gruman \& Melrose 2002).

\medskip
\noindent
{\it Other maser mechanisms}:
The other three instability-based mechanisms are linear acceleration emission (Melrose 1978; Rowe 1995), anomalous Doppler instability (Machabeli \& Usov 1979) and curvature-drift instability (Kazbegi, Machabeli \& Melikidze 1991). Linear acceleration emission requires large-amplitude oscillations in the parallel electric field as the driving mechanism in the source region. Although there is no strong argument against this mechanism, there is no detailed model for the development of the required large-amplitude electrostatic oscillations and no independent evidence for them. The other two mechanisms are possible only in the outer magnetosphere (Lyutikov, Blandford \& Machabeli 1999), and are sensitive to $B$, both of which imply that they cannot apply to all classes of pulsars. Also, both require that the refractive index be greater than unity at the point of emission, and this is even more difficult to satisfy than for the beam-driven Cerenkov instability.

\medskip
\noindent
{\it Orthogonal modes}:
There is compelling evidence that in at least some pulsars the emission emerges in two orthogonal modes (OMs) that become spatially separated (e.g., McKinnon \& Stinebring 2000). The presence of OMs requires refraction in a birefringent plasma that causes different rays for the two OMs. The observed OMs raise a further difficulty: any instability favors one wave mode in the plasma over others, and after many growth times the faster growing mode completely dominates, implying emission in only one of the natural modes of the plasma.  Hence, one requires some process that converts waves initially purely in one mode into a roughly equal mixture of two modes. Mode coupling due to large-scale gradients is one possibility, but a more effective process such as reflection off sharp boundaries seems to be necessary. Moreover, if such propagation effects are important, then the observed polarization should be determined at a polarization limiting region, rather than being characteristic of the emission mechanism or of the dominant mode at the source.

\section{Source regions and heights of emission}

It has long been thought that the most plausible location for the source region of the radio emission is many tens of radii above the star in the polar cap regions. Recent analysis of data (Gangadhara \& Gupta 2001) favors a source region at such heights, roughly midway between the magnetic axis and the last closed field lines, far from the gap regions where pair creation is conventionally assumed to occur. It may be that significant pair creation does occur between the conventional inner and outer gaps. However, why should the radio emission originate from intermediate heights, rather than the gaps where the pair creation should be most efficient?

A speculation as to how some of these difficulties might be overcome involves appealing to a possible analogy with solar radio bursts. Type~III bursts are due to a stream of electrons, that generate Langmuir waves through a Cerenkov instability, and types~I, II and~III are all attributed to plasma emission. Theory implies that the polarization of plasma emission should be 100\% in the o~mode, and this is sometimes the case for type~I bursts, but never the case for type~II and~III bursts, and some bursts of all types can be unpolarized. A depolarizing agent operates, and this requires very small-scale inhomogeneities (Melrose 1975), and may be associated with reflection off  overdense structures, such as columns invoked to account for the directivity of type~I emission (Bougeret \& Steinberg 1977). Ducting is also required to account for the apparent sources being scatter images at much greater heights than the actual sources (Duncan 1979). The polarization, directivity and apparent height can all be explained in terms of a source in an under\-dense region with the escaping radiation reflected off nearly radial overdense structures.

Nonstationary, inhomogeneous pair creation would produce dense columns of pair plasma that can act as ducts for radio emission generated in the under\-dense regions between the columns. Reflection off the columns can convert radiation in one mode into a mixture of the two modes, as well as ducting the radiation outward. The radiation characteristic would then correspond to those of an apparent source at a much greater height than the actual source. These speculations suggest that an appeal to locally highly time-variable pair creation may resolve some of the outstanding difficulties in the interpretation of the source  location and polarization of pulsar radio emission.

\section{Conclusions} 

High-energy emission processes are well understood, but the location of the source region is uncertain. The radio mechanism is probably either relativistic plasma emission or torsion-driven curvature maser instability; a detailed critical comparison of these is needed. Observed orthogonal mode polarization appears to require structures with sharp boundaries off which the waves can be reflected.

\end{document}